# Band Alignment of Grafted versus Directly Bonded Si/Ge Heterostructures

Pargam Vashishtha[1], Yang Liu[1], Tsung-Han Tsai[1], Yi Lu[1], Samuel A. Haessly[1], Chenyu Wang[1], Nanditha M. Dissanayake[2], Edward, K. Huang[2], Zhenqiang Ma[1,*]

[1]*Department of Electrical and Computer Engineering, University of Wisconsin-Madison, Madison, Wisconsin 53706, USA.*

[2]*Attollo Engineering, Camarillo, CA 93012, USA*

*Corresponding Author's Email: mazq@engr.wisc.edu*

## Abstract

The electronic properties of semiconductor heterostructures are influenced by the local bonding and electrostatic environment at the heterointerface. Here, we investigate how the interface-formation mechanism modifies the band alignment of Si/Ge heterostructures by directly comparing $Al_2O_3$-mediated semiconductor grafting with direct bonding. Interface-sensitive X-ray photoelectron spectroscopy combined with Kraut analysis reveals type-I band alignment for both structures, but with different band offsets: the valence-band offset increases from ~0.08 eV for directly bonded Si/Ge to ~0.21 eV for grafted Si/$Al_2O_3$/Ge, accompanied by a corresponding reduction in the conduction-band offset. Raman spectroscopy further distinguishes the two interfacial configurations. The different core-level shifts and band offsets indicate that the electronic structure is sensitive to the interfacial bonding and electrostatic environment introduced by the two integration schemes. These results demonstrate that, even for an identical Si/Ge material pair, the interface-formation pathway provides an additional degree of freedom for controlling heterojunction band alignment, offering a route toward interface-engineered electronic and optoelectronic devices.

Semiconductor heterostructures enable the integration of materials with complementary electronic properties and are fundamental to advanced electronic and optoelectronic devices.[1] Their performance is strongly governed by the electronic structure at the heterointerface, where band alignment influences carrier injection, transport, confinement, and leakage.[2] In addition to the intrinsic properties of the constituent materials, interfacial bonding and charge redistribution can modify the local electrostatic potential and consequently the band alignment. Therefore, understanding and controlling the heterointerface is essential for achieving high-performance heterogeneous semiconductor devices.

Among semiconductor heterostructures, silicon (Si)/ germanium (Ge) is attractive because it combines the mature processing and integration capabilities of Si with the high carrier mobility and narrow bandgap of Ge, making it promising for high-speed electronic and near-infrared optoelectronic devices. However, the lattice mismatch between Si and Ge still presents a major challenge for conventional heteroepitaxial integration, often generating misfit and threading dislocations that can degrade junction performance. Buffer-layer engineering and high-temperature treatments can reduce these defects but increase process complexity and thermal budget and may promote Si-Ge interdiffusion.[3-6] These limitations motivate alternative approaches for integrating Si and Ge while maintaining a high-quality heterointerface.

Bonding (i.e., wafer fusion)-based integration provides an alternative by joining independently prepared Si and Ge without heteroepitaxial growth across the heterojunction. Direct wafer and nanomembrane bonding have demonstrated Si/Ge integration at relatively low processing temperatures.[7, 8] Bonded Ge/Si heterojunction diodes and photodetectors were demonstrated previously.[9, 10] However, the electrical characteristics of bonded Si/Ge junctions remain highly sensitive to surface preparation, residual interfacial layers, interface states, and trapped charges, which can affect the interfacial potential distribution and carrier transport.[11, 12] Effective interface passivation is therefore critical for obtaining high-quality rectifying Si/Ge heterojunctions.[11] Semiconductor grafting provides an alternative interface-engineering approach by incorporating an ultrathin dielectric-mediated interface that facilitates heterogeneous integration while providing double-side interface passivation.[13-21] Importantly, the dielectric interlayer also introduces additional engineering parameters, including its material, thickness, interfacial bonding, and processing conditions, that may be used to tailor the heterojunction electronic structure.

Despite the excellent electrical characteristics demonstrated for grafted Si/Ge heterojunctions[13], which outperform all other Si/Ge heterojunctions formed via various other methods, a fundamental understanding of their interfacial bonding properties and resulting band alignment remains lacking. Understanding this relationship is essential because the interfacial bonding configuration can influence charge redistribution and the local electrostatic environment, thereby modifying band offsets and carrier-transport barriers.

Although grafted Si/Ge heterojunctions exhibit unprecedented electrical characteristics, direct bonding can be adopted more easily in industrial fabrication at present.[7, 8] Therefore, comparing the band alignment between the grafted and the directly bonded Si/Ge heterojunctions is pervasive and momentous, as both heterojunctions are being optimized to improve their performance but under completely different interface bonding conditions. Comparing these interfaces will also provide insight into how the grafting-mediated interface influences the Si/Ge electronic structure. Here, we systematically compare $Al_2O_3$-mediated grafted and directly bonded Si/Ge heterostructures using X-ray photoelectron spectroscopy and Kraut band-offset analysis. The different band offsets demonstrate the critical role of interface formation in determining the Si/Ge electronic structure and highlight semiconductor grafting as a promising strategy for simultaneously engineering interface passivation and band alignment.

The grafted Si/Ge heterostructure, as shown in Fig. 1 (a), consists of a 205 nm-thick single-crystalline Si, an $Al_2O_3$ layer deposited by atomic layer deposition (ALD) for 5 cycles, and a high-quality 1 μm (top 100 nm highly doped $1\times10^{18}$ $cm^{-3}$ and 900 nm unintentionally doped) thick Ge layer. The Si layer was released from a 205 nm/400 nm silicon-on-insulator (SOI) that was fabricated by Soitec®. The high-quality Ge layer was grown in an Epsilon® chemical vapor deposition (CVD) system at Advanced Epi Materials and Devices Ltd. The layer structures of the Si and Ge source materials, together with the crystalline quality assessment of the Ge thin film (Fig. S1) and the fabrication process of the grafted Si/Ge (i.e., Si/$Al_2O_3$/Ge) structures (Fig. S2), can be found in the Supporting Information (SI). An optical microscopy image of the grafted Si/$Al_2O_3$/Ge heterostructure is shown in Fig. S2 ($c_1$).

The directly bonded Si/Ge heterostructures consist only of 205 nm Si and Ge (i.e., Si/Ge), without the incorporation of an ALD-deposited $Al_2O_3$ interlayer at the interface. The interface was kept free of impurities, especially native oxide, as described below. To achieve this impurity-free Si-Ge interface, the Si/Ge heterostructure was fabricated using a transfer-and-undercut bonding process. The detailed process is shown in Fig. S3, SI. Briefly, we first transferred a Si nanomembrane (NM) onto a Ge substrate coated with a ~400 nm $SiO_2$ layer, then annealed to promote strong bonding. The $SiO_2$ layer was then removed by sacrificial HF etching, followed by rinsing. During the entire etching and rinsing process, the sample was immersed in the liquid without air exposure. After drying, the released Si was weakly and intimately bonded to the Ge substrate via van der Waals forces. The Si/Ge heterostructure was completed by annealing under the typical conditions used for wafer bonding.[22] Although the direct bonding method in this work is different from the conventional vacuum-based wafer bonding method, the interface of the directly bonded structure in this work is expected to resemble that of the wafer-bonded structure closely.[7] The schematic and an optical microscopy image of the directly bonded Si/Ge heterostructure are shown in Fig. 1(b) and Fig. S3 (e), SI, respectively. It is noted that (identical) thin single-crystalline Si NMs were used in both heterostructures instead of a rigid Si wafer. Using Si NMs does not alter the underlying interface bonding chemistry in either heterostructure but greatly eases X-ray access to the interface for the following analyses.

Raman spectroscopy was performed on the two heterostructures immediately after they were fabricated and prior to Si NM thinning, and the resulting spectra are shown in Fig. 1(c) together for comparison, along with a Raman spectrum of pristine Ge. In addition to the dominant Ge-Ge optical phonon mode near ~300 $cm^{-1}$, all three samples exhibit broad low-intensity second-order Ge-related Raman features in the ~350-380 $cm^{-1}$ region.[23] The deconvoluted Raman spectra of these features for the three samples are shown in Fig. 1(d), Fig. 1(e), and Fig. 1(f), respectively. Notably, the directly bonded sample exhibits a distinct broad feature around ~400 $cm^{-1}$ (Fig. 1(e)) that is absent in the other samples.[24] The unique peak around 400 $cm^{-1}$ (Fig. 1(e)) is attributed to the vibrational signature of Si-Ge bonds,[25] thereby confirming the formation of direct bonding in the Si/Ge heterostructure. The absence of the broad feature around ~400 $cm^{-1}$ for the Si/$Al_2O_3$/Ge sample (Fig. 1(d)) indicates that the top Si layer and the bottom Ge are physically separated from each other by the ALD $Al_2O_3$, a distinct feature of all grafted heterostructures. [17]

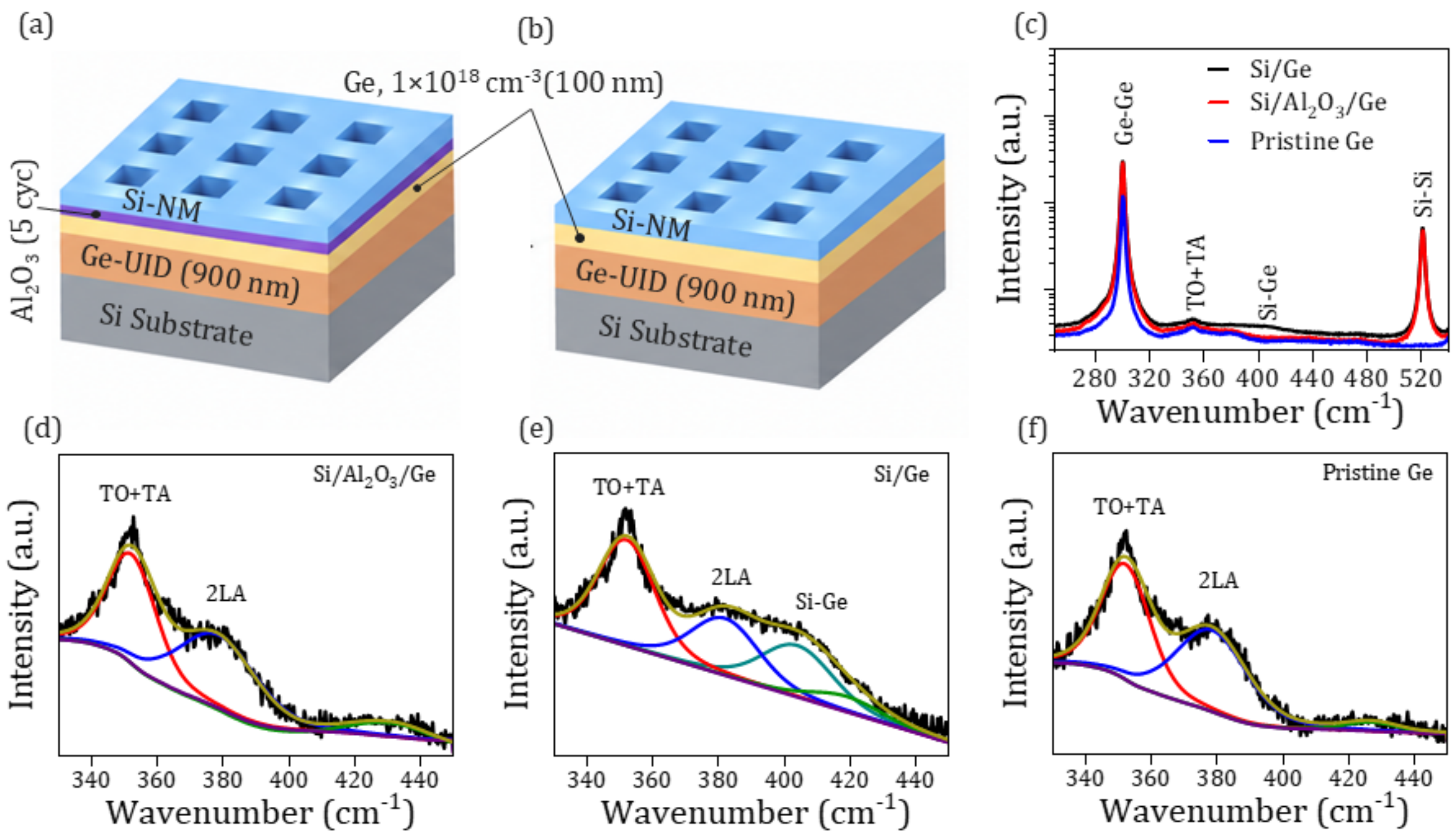

***Figure 1.*** *Schematic illustration of the two heterostructures. (a) Grafted Si/Ge heterostructure (Si/$Al_2O_3$/Ge). (b) Directly bonded Si/Ge heterostructure. Fabrication process details and illustrations can be found in the SI. (c) Raman spectra of the Si/$Al_2O_3$/Ge (red) and the Si/Ge (black) heterostructures, along with the Raman spectrum of a pristine Ge substrate (blue). (d), (e) and (f). Detailed deconvoluted Raman spectra including the transverse acoustic (TA), transverse optical (TO), and longitudinal acoustic (LA) phonon contributions in the Si/$Al_2O_3$/Ge and Si/Ge heterostructures, together with pristine Ge for comparison, respectively.*

X-ray photoelectron spectroscopy (XPS; Thermo Scientific K-alpha) was used to analyze the chemical and electronic structure of the two heterointerfaces. Calibration was performed using the Au $4f_{7/2}$ peak at 84.0 eV. The spectra were recorded with a pass energy of 10 eV, an analysis spot size of 400 µm, a step increment of 0.02 eV, and a maximum dwell time of 10 Sec. All XPS spectra were analyzed using CasaXPS with Shirley background subtraction.

To apply Kraut's method for construction of band alignment, XPS spectra of Ge 3d and Si 2p were obtained from pristine Ge and pristine Si NM samples.[26] The pristine Si NM and Ge were cleaned in situ with a 200 eV $Ar^+$ ion beam to remove native oxides and carbon contamination before taking the spectra. Fig. 2(a) and Fig. 2(b) show the XPS core-level spectra acquired from the pristine Si NM and the pristine Ge, respectively. The Si $2p_{3/2}$ core-level peak appears at 99.46 eV for pristine Si NM, and the Ge $3d_{5/2}$ peak appears at 29.27 eV for pristine Ge. In addition, the valence band maxima (VBM) for pristine Si NM (Fig. 2(c)) and Ge (Fig. 2(d)) were extracted from the valence band spectra near the Fermi level by linear extrapolation of the leading edge of the density of states. The extracted VBM values are 0.67 eV for pristine Si-NM and 0.32 eV for pristine Ge.

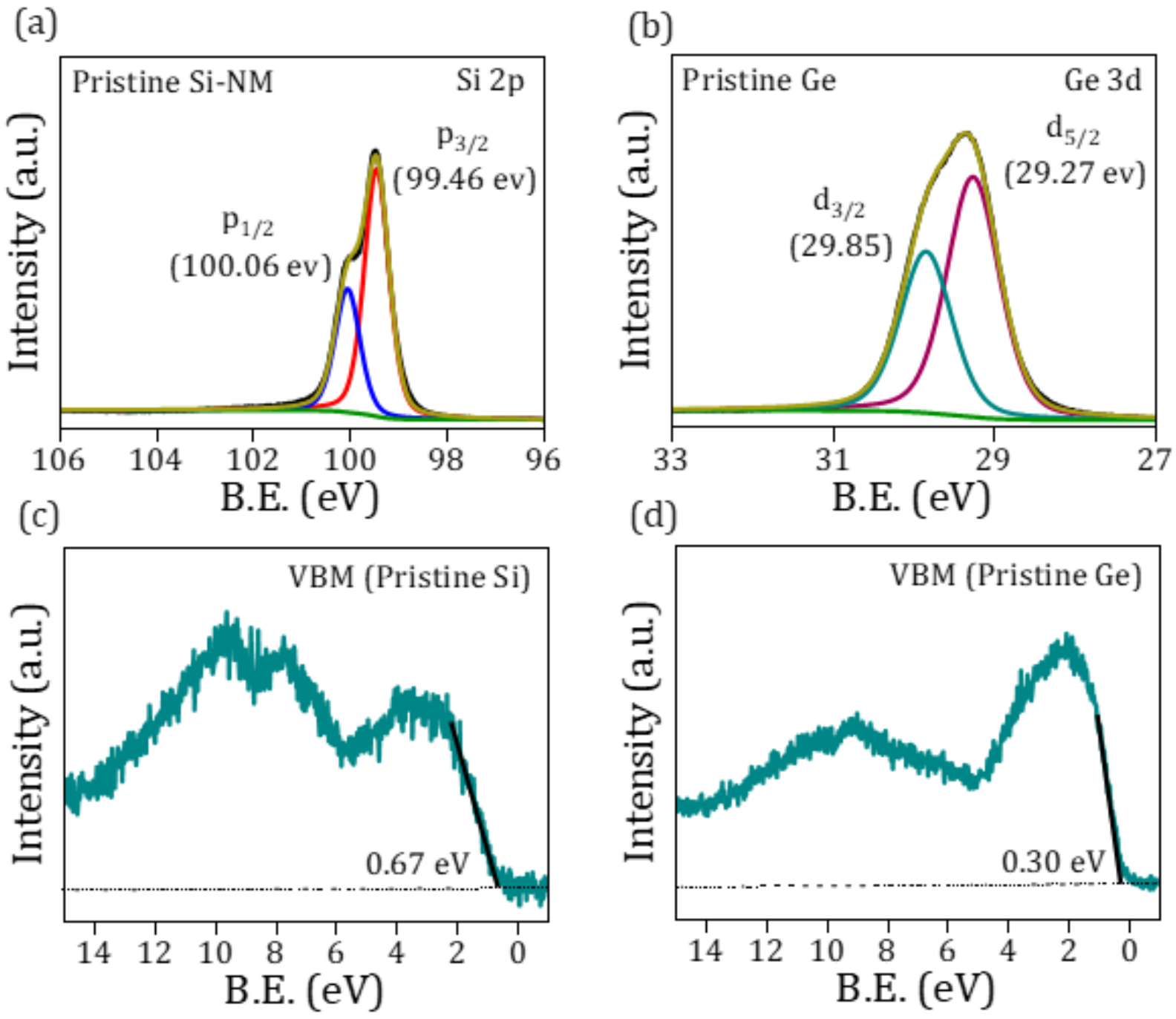

***Figure 2.*** *XPS core-level spectra acquired from the pristine Ge and Si samples. (a) Si 2p spectrum collected from the pristine Si NM. (b) Ge 3d spectrum recorded from the pristine Ge sample. Valence-band spectra obtained from the (c) pristine Si NM surface and (d) pristine Ge surface.*

To access the interfaces of the two heterostructures, the Si layers in the two heterostructures (Fig. 1 (a) and Fig. 1(b)) were thinned down using reactive ion etching with $CF_4/O_2$, followed by controlled in situ $Ar^+$ ion-beam etching in the XPS chamber (2 mm raster size, 200/500 eV beam energy, monatomic mode). Briefly, the sample was transferred to the XPS chamber under nitrogen to minimize surface oxidation during each thinning step. Within the XPS system, in situ $Ar^+$ ion-beam etching was performed progressively to approach the two interfaces. As etching advanced, the Si 2p peak intensity systematically decreased, while the Ge 3d peak intensity increased, confirming successful profiling toward the interfacial region (See Fig. S5 & Fig. S6, SI). When the Si 2p and Ge 3d peak intensities became comparable, the analysis region was near the interface. The estimated thickness of the remaining Si should be less than 10 nm. The stable peak position and consistent line shape of the Si 2p (Fig. S5, SI) feature throughout the etching sequence confirm that the single-crystalline quality of the Si NM remained intact during deep ion milling, thereby validating the top-down XPS interface-profiling approach. Si thinning is complete when both Si and Ge signals are clearly resolved. It is noted that the two heterostructure samples were analyzed at two different times but under identical conditions. Following the completion of the top Si thinning process, high-resolution core-level XPS spectra of Ge 3d and Si 2p at the heterointerface of the two heterostructures were acquired at a fixed pass energy.

Fig. 3(a) and Fig. 3(b) show the XPS core-level spectra acquired from the interface of the Si/$Al_2O_3$/Ge heterostructure. The Si $2p_{3/2}$ core-level peak is observed at 99.51 eV for the heterostructure (Fig. 3(a)), and the Ge $3d_{5/2}$ peak appears at 29.48 eV (Fig. 3(b)). Fig. 3(c) and Fig. 3(d) show the XPS core-level spectra acquired from the interface of the Si/Ge heterostructure. The Si $2p_{3/2}$ core-level peak is observed at 99.54 eV, and the Ge $3d_{5/2}$ peak appears at 29.64 eV.

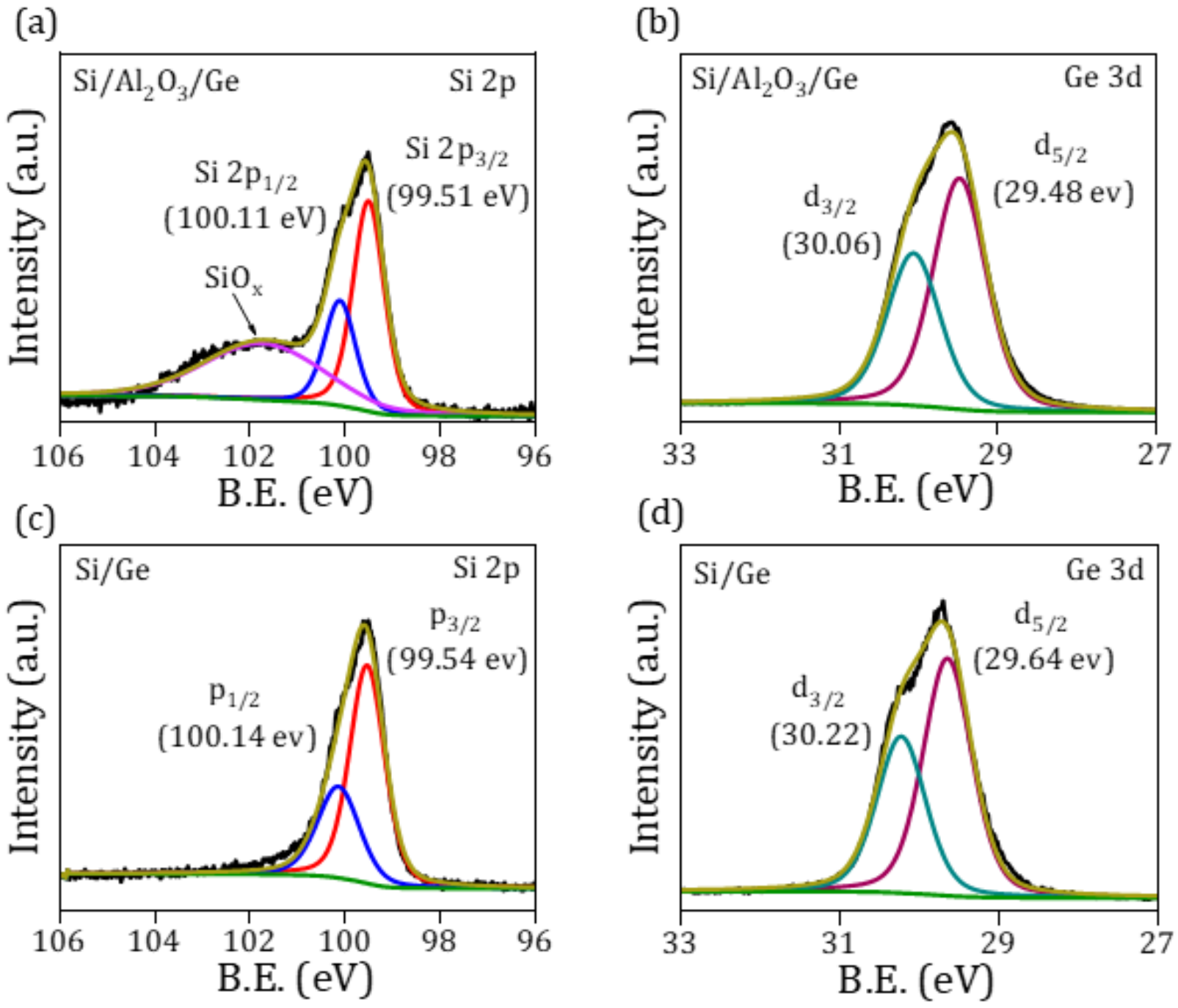

***Figure 3.*** *XPS core-level spectra acquired from the interface of the Si/$Al_2O_3$/Ge and the Si/Ge heterostructures. (a) Si 2p spectrum and (b) Ge 3d spectrum measured at the heterointerface of the Si/$Al_2O_3$/Ge heterostructure. (c) Si 2p spectrum and (d) Ge 3d spectrum measured at the heterointerface of the Si/Ge heterostructure.*

The Si and Ge core-level peak positions at the Si/Ge interface shift to higher binding energies than those at the Si/$Al_2O_3$/Ge interface. This shift is associated with interfacial charge redistribution and the formation of covalent Si-Ge bonds, which modify the local electrostatic potential of Ge atoms at the interface. Direct bonding alters the initial-state electronic structure, leading to a chemical shift in the Ge core level. This behavior highlights the strong influence of oxide-free interfacial bonding on the electronic states of Ge and underscores the role of interfacial chemistry.

Valence-band offsets for both heterostructures were calculated using the Kraut method:[26]

$$\Delta E_V = (E_{CL} - E_{VBM})_{Si} - (E_{CL} - E_{VBM})_{Ge} + \left(E_{CL,Si} - E_{CL,Ge}\right)_{Interface}$$

where $E_{CL}$ and $E_{VBM}$ represent the core levels of materials and the valence-band maximum, respectively. All the core-level spectrum and valence-band parameters used for band-offset determination are summarized in Table 1. A Ge bandgap of 0.66 eV was adopted from the literature for the band-alignment calculations. The corresponding band diagrams for the $Si/Al_2O_3/Ge$ and the Si/Ge heterostructures are shown in Fig. 4 (a) and 4(b), respectively.

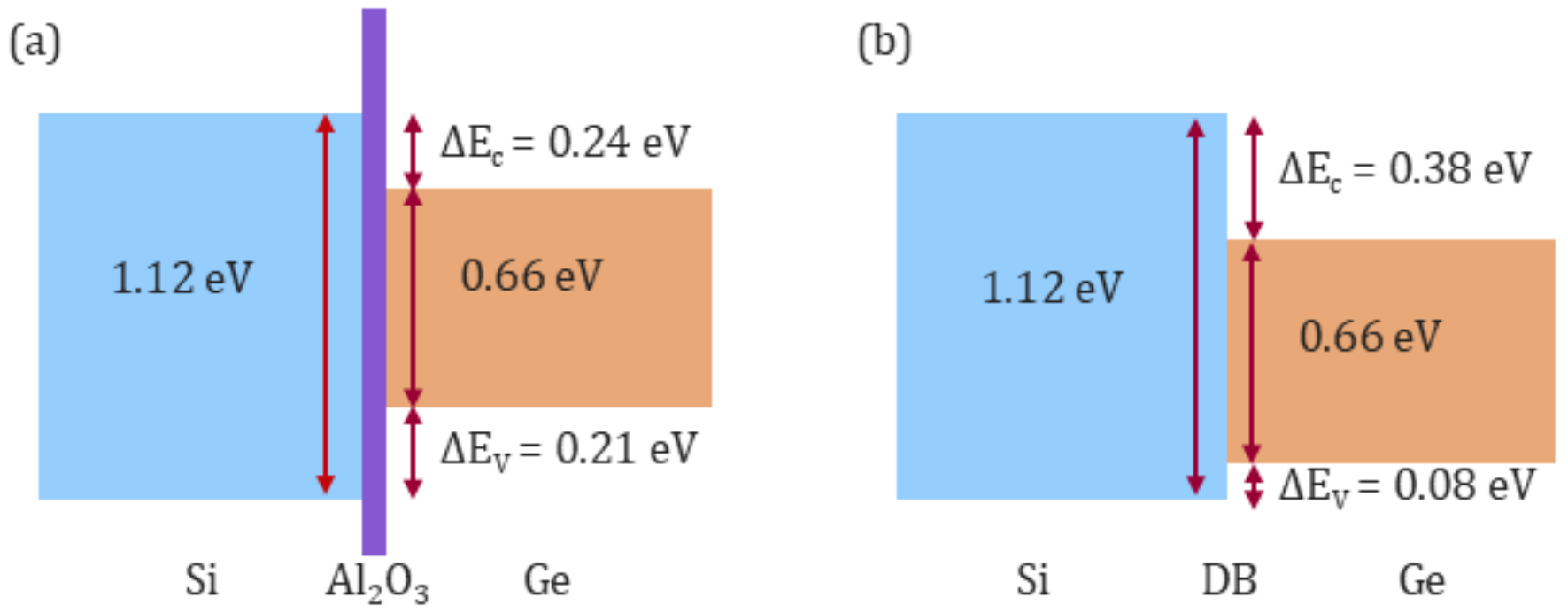

***Figure 4.*** *Band structure constructed for the (a) $Si/Al_2O_3/Ge$ and (b) Si/Ge heterostructures.*

***Table 1.*** *Compiled Ge 3d and Si 2p core-level binding energies with the corresponding VBM positions for pristine Si-NM, pristine Ge, $Si/Al_2O_3/Ge$, and Si/Ge heterostructures, along with the values of calculated valence- and conduction-band offsets.*

| ***Core-level*** / ***Material*** | *VBM* | *Si 2p$_{3/2}$ ($E_{CL,\ Si}$)* | *Ge 3d$_{5/2}$ ($E_{CL,\ Ge}$)* | *VBO ($\Delta E_V$)* | *CBO ($\Delta E_C$)* |
|---|---|---|---|---|---|
| *Pristine Ge* | *0.67* | *99.46* | - | - | - |
| *Pristine Si-NM* | *0.30* | - | *29.27* | - | - |
| *$Si/Al_2O_3/Ge$* | - | *99.51* | *29.48* | *0.21* | *0.24* |
| *Si/Ge* | - | *99.54* | *29.64* | *0.08* | *0.38* |

Both heterostructures exhibit type-I band alignment. The Si/Ge heterostructure exhibits a valence band offset (VBO or $\Delta E_V$) value of 0.08 eV. The $Si/Al_2O_3/Ge$ heterostructure exhibits a larger VBO of 0.21 eV. The difference of VBO values indicates that the interface formation mechanisms are different, which is also evidenced in Fig. 1(d) and Fig. 1(e). It is noted that both band alignments deviate from one (a "utopian" type-II) that can be easily constructed using Anderson's electron affinity rule (Si and Ge electron affinity energy values are 4.05 eV and 4.0 eV, respectively). The deviation of the Si/Ge heterostructure band alignment from the utopian one roots from the fact that the real Si-Ge bonds are far different from the utopian interface where Si and Ge would have a lattice-matched ideal bonding. The distorted Si and Ge lattice at the Si/Ge heterointerface results in possible dangling bonds and trap states, and therefore local charge creation, which is reflected in the measured band alignment.

In the Si/$Al_2O_3$/Ge heterostructure, the ultrathin $Al_2O_3$ layer was employed to provide double-side interface passivation, such that the dangling bonds of the Si surface and the Ge surface are passivated through the formation of Si-O (See Fig. 3 (a)).[13] The Ge-O bonds were not seen in Fig. 3(b) because oxygen tends to bond with Si rather than Ge.[11] The $Al_2O_3$ physically separated the Si and Ge (but allows quantum tunneling). Thus, the traps induced by the imperfect Si-Ge bonding at the interface of the Si/Ge heterostructure have been greatly reduced. The effects of the remaining traps due to incomplete Si-O and Ge-O bonds on the band alignment are expected to be minimal. However, $Al_2O_3$ induces its own charge due to the amorphous state and non-stoichiometry.[27] In addition, the Si-O and Al-O bonds also introduce dipoles at the interface due to the different electronegativities of Si, O, and Al. Charge redistribution at the local bonding environment leads to the formation of interface dipoles and an associated electrostatic potential, consequently exerts its influence on the overall band alignment of the Si/$Al_2O_3$/Ge heterostructure.[28] The degree of influence may vary depending on processing parameters, including the type of the interface ultrathin dielectric, the number of ALD cycles of the ultrathin dielectric, the annealing temperature, and annealing time for the formation of chemical bonds at the interface. The conditions adopted in this study, identical to our previous work (5 cycles of ALD $Al_2O_3$ at 200°C for 5 min), were chosen to optimize the current-voltage characteristics of the grafted Ge/Si *p-n* diodes.[13]

It is noted that in comparison to the two band alignments that are shown in Fig. 4, the band alignment of Si/Ge is more favorable for a *p*-Ge/*n*-Si diode than that of the Si/$Al_2O_3$/Ge. However, the previous experiment showed that the grafted *p*-Ge/*n*-Si diode (ideality factor n = 1.04, $I_{on}/I_{off}$=1.5 × $10^8$ at +/- 1 V) showed a much better performance than the directly bonded *p*-Ge/*n*-Si diode (n = 2.28, $I_{on}/I_{off}$= 8 × $10^3$ at +/- 1 V).[17, 8] The relatively lower performance of the directly bonded *p*-Ge/*n*-Si diode indicates that its interface is not as well-passivated as the grafted *p*-Ge/*n*-Si diode interface and that the band alignment of the directly bonded heterostructure is not the only metric to predict the electronic characteristics of a related *p-n* junction. On the other hand, since the interface of a grafted Si/Ge heterostructure (i.e., Si/$Al_2O_3$/Ge) is well passivated, the band alignment becomes a useful metric for predicting the electronic characteristics of a related p-n junction.

Importantly and comparatively, direct bonding offers fewer parameters for controlling the interfacial electronic structure and lacks such dielectric-mediated double-side passivation as in the grafted heterostructure. Semiconductor grafting therefore provides a more versatile platform for Si/Ge band-alignment engineering. Systematic investigation of these parameters may enable detailed control of the Si/Ge interfacial electronic structure.

This work demonstrates that the band alignment of Si/Ge heterostructures is not determined solely by the semiconductor materials (e.g., electron affinity values) but is strongly influenced by how the heterointerface is formed. By comparing $Al_2O_3$-mediated grafted and directly bonded Si/Ge structures prepared from the same semiconductor pair, interface-sensitive XPS and Kraut analysis reveal type-I alignment in both cases, with a VBO of ~0.08 eV for directly bonded Si/Ge and ~0.21 eV for grafted Si/$Al_2O_3$/Ge. The corresponding CBO changes from ~0.38 eV to ~0.25 eV, demonstrating a substantial redistribution of the electron and hole barriers produced by the different interface configurations.

The spectroscopic results further indicate distinct interfacial environments: the directly bonded structure exhibits the ~400 $cm^{-1}$ Raman feature associated with Si-Ge bonding, whereas the grafted structure contains an ultrathin $Al_2O_3$-mediated interface with Si-O -related chemistry and additional dielectric-associated electrostatic contributions. These differences provide a physical basis for the observed band-offset variation, while the XPS core-level shifts themselves are interpreted as reflecting the combined influence of local bonding and interface electrostatics.

Overall, the results establish the interface-formation pathway as an experimentally accessible parameter for tuning band alignment in lattice-mismatched heterostructures and provide a basis for designing grafted interfaces with tailored carrier-injection and confinement barriers.

## Acknowledgements:

This research was sponsored by Air Force Research Lab through Attollo Engineering under the grant number FA237724CB032.

## Conflict of Interest:

The authors declare no conflict of interest.

## Data Availability Statement

The data that support the findings of this study are available from the corresponding author upon reasonable request.

# Supporting Information

## Band Alignment of Grafted versus Directly Bonded Si/Ge Heterostructures

Pargam Vashishtha[1], Yang Liu[1], Tsung-Han Tsai[1], Yi Lu[1], Samuel A. Haessly[1], Chenyu Wang[1], Nanditha M. Dissanayake[2], Edward, K. Huang[2], Zhenqiang Ma[1,*]

[1]*Department of Electrical and Computer Engineering, University of Wisconsin-Madison, Madison, Wisconsin 53706, USA.*

[2]*Attollo Engineering, Camarillo, CA 93012, USA*

**Corresponding Author's Email: mazq@engr.wisc.edu*

Fig. S1 summarizes the structures of the starting pristine Si and Ge materials and the structural characterization of the Ge epilayer thin film. Fig. S1(a) and S1(b) show schematic vertical layer stacks of a silicon-on-insulator (SOI) wafer and a Ge epilayer, respectively, identifying the individual layers used for heterostructure fabrication. Moreover, to evaluate the crystalline quality and strain relaxation of the chemical vapor deposition (CVD)-grown pristine Ge thin film, high-resolution X-ray diffraction (HR-XRD) 2θ/ω scans were performed around the (004) reflection, as shown in Fig. S1(c). The diffraction pattern exhibits two intense peaks at 66.07° and 69.12°, corresponding to the Ge (004) and Si (004) crystalline substrates, respectively. The sharp and well-defined Ge peak at 66.07° indicates high crystalline quality and a largely strain-relaxed Ge film.[1] A slight asymmetry on the high-angle side of the Ge (004) peak suggests the presence of residual strain and/or compositional variation near the Ge/Si interface. To further assess the structural quality of the film, reciprocal space mapping (RSM) was performed around the asymmetric (224) reflection, as shown in Fig. S1(d). Two distinct diffraction peaks corresponding to the Ge layer and the Si substrate are observed. The Ge peak lies close to the relaxation line, indicating that the Ge layer is ~99.87% relaxed with a residual strain of ~0.17%.[1] Atomic force microscope (AFM) measurements over a 10 × 10 $\mu m^2$ scan area (Fig. S1(e)) reveal a root-mean-square surface roughness of 2.37 nm. Collectively, these

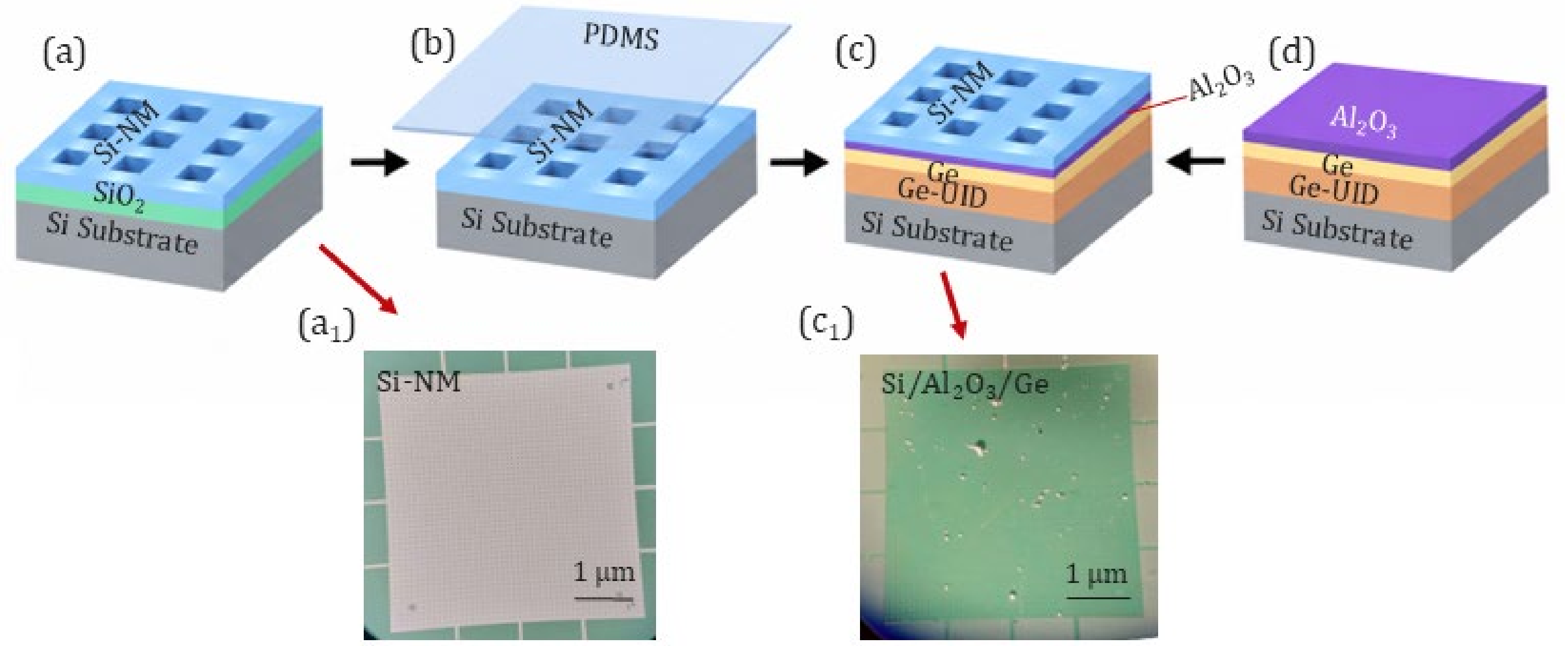

***Figure S2.*** *(a) Schematic depicting the fabrication of the Si/$Al_2O_3$/Ge heterostructure. Together with optical microscopy images of the released Si-NM (S2($a_1$)) and the completed Si/$Al_2O_3$/Ge heterostructure (S2($c_1$)).*

A direct-bonded Si/Ge heterostructure was fabricated using a transfer-and-undercut bonding approach, as illustrated step-by-step in Fig. S3. The Si-NM was first released from the SOI wafer and transferred onto the 400 nm $SiO_2$-coated Ge thin film, followed by annealing at 250°C for 5 min to establish initial bonding. The interfacial $SiO_2$ layer was then selectively removed by repeated immersion in an HF/HCl (3:5) solution, followed by rinsing in IPA/ACT (1:1) to remove residual etchants. This undercut process enabled direct contact between the Si-NM and Ge without exposing the bonding interface to air. After complete removal of the interfacial $SiO_2$, the directly bonded Si/Ge heterostructure was annealed at 400°C for 60 min to enhance its mechanical and chemical stability.[4] This approach avoids native oxide formation at the Si/Ge interface, which is typically prevented by employing a vacuum environment in conventional Si/Ge direct-bonding processes.

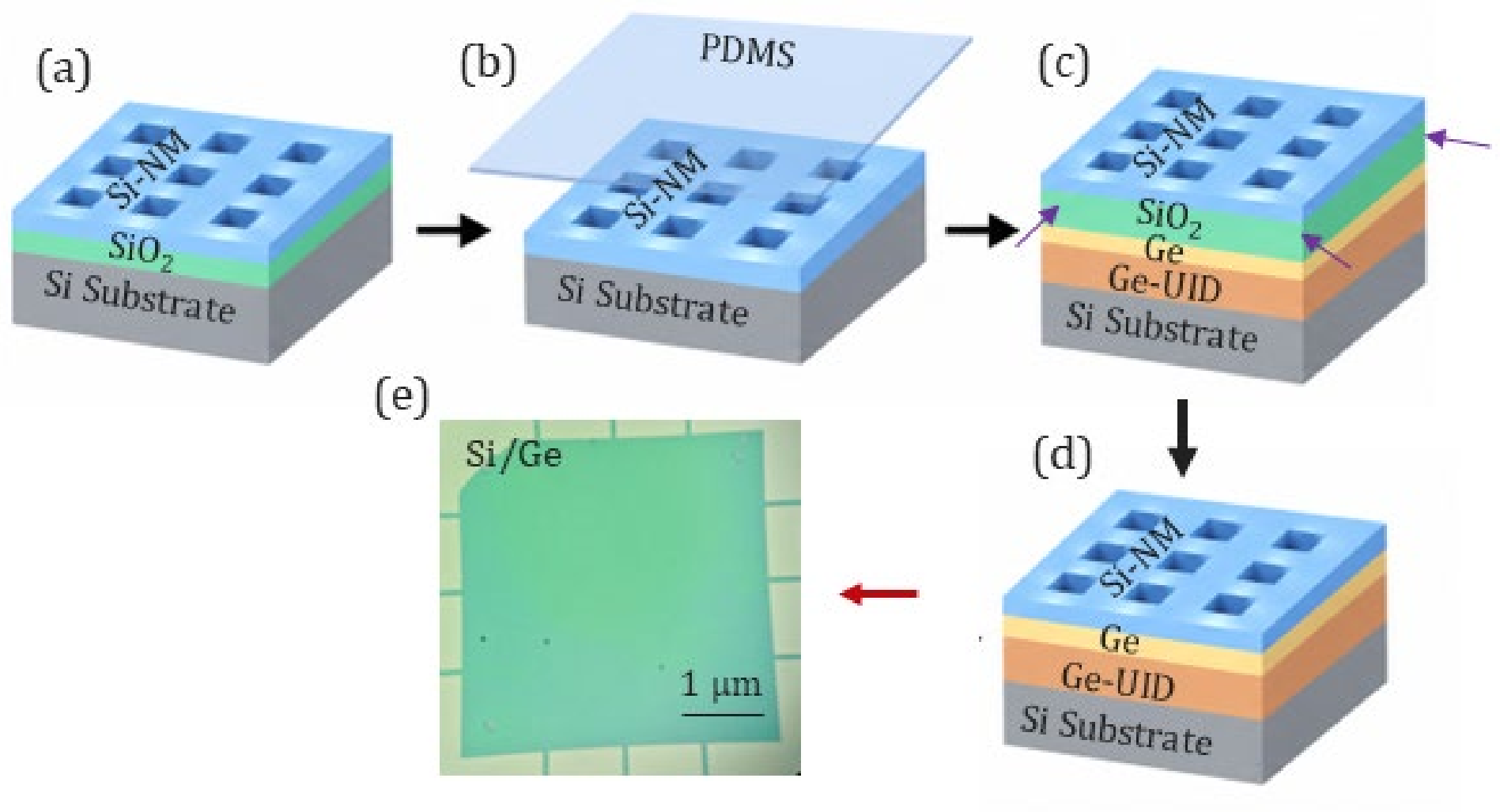

***Figure S3.*** *(a) Schematic depicting the fabrication of the direct-bonded Si/Ge heterostructure.*

High-resolution X-ray photoelectron spectroscopy (XPS) core-level spectra acquired from the bare Ge and Si-NM are shown in Fig. S4. The C 1s and O 1s spectra of the bare Ge substrate, displayed in Fig. S4(a) and S4(b), indicate effective removal of surface carbon contamination and native oxide following cleaning. Similarly, the C 1s and O 1s spectra obtained from the bare Si nanomembrane, shown in Figures S4(c) and S4(d), exhibit negligible residual contamination.

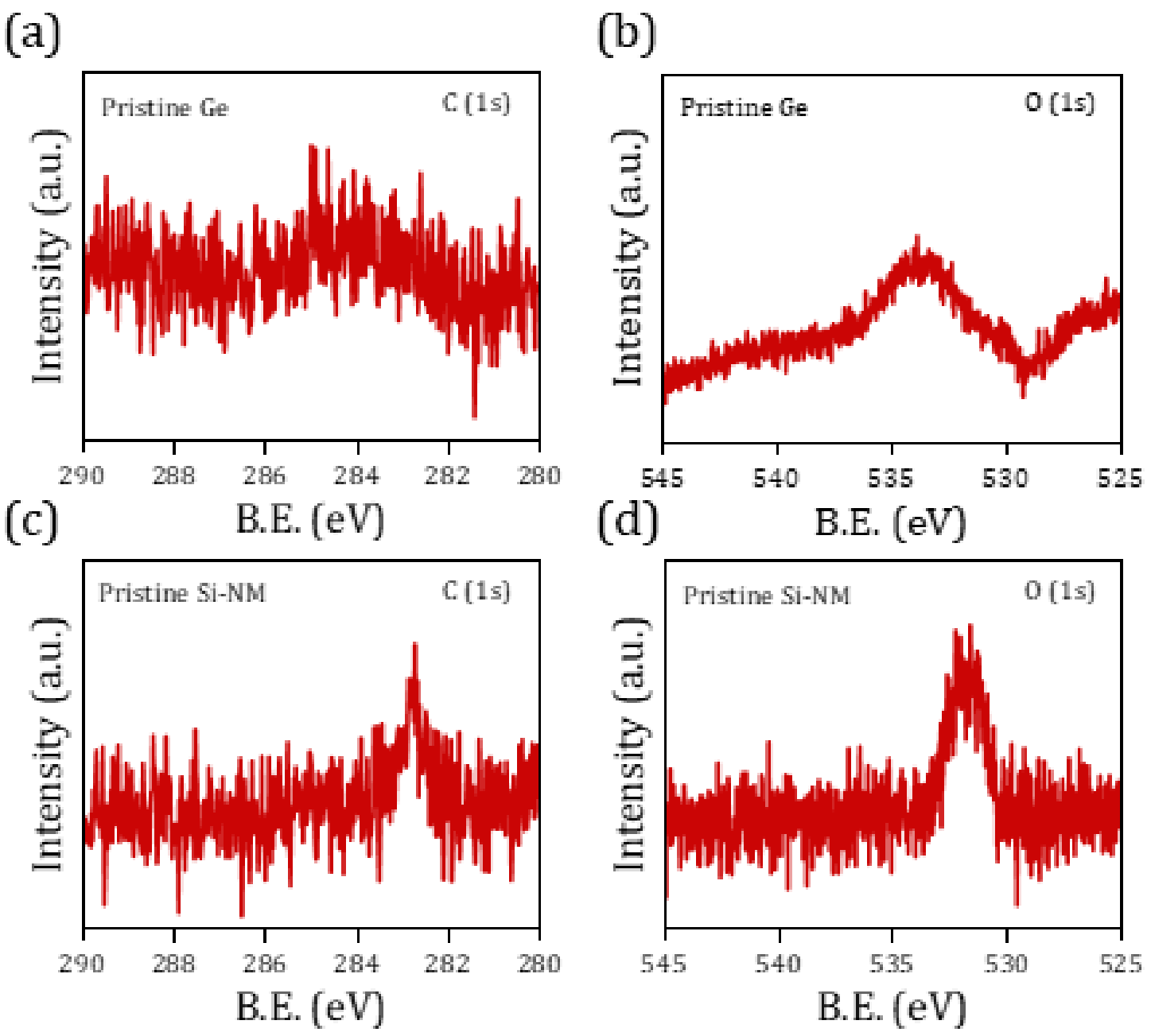

***Figure S4.*** *XPS core-level spectra of (a) C 1s, (b) O 1s for bare Ge and (c) C 1s, and (d) O 1s for bare Si-NM.*

Fig. S5 illustrates the evolution of XPS signals during stepwise in-situ $Ar^+$ ion etching used to access the buried $Si/Al_2O_3/Ge$ interface. Fig. S5(a) shows the etching profile as a function of etching time and gun energy, demonstrating controlled thinning of the Si-NM in two energy steps: 200 eV and 500 eV. Fig. S5(b) and S5(c) present the corresponding Ge 3d and Si 2p core-level spectra acquired at different etching stages. As etching progresses, the Ge 3d signal gradually increases while the Si 2p intensity decreases, confirming proximity to the interface. Fig. S5(d) quantitatively summarises the intensity evolution of the Ge 3d and Si 2p peaks with etching time, clearly indicating the approach to the interface. Moreover, the sudden change in the Ge intensity after 90 sec is due to a change in the ion energy.

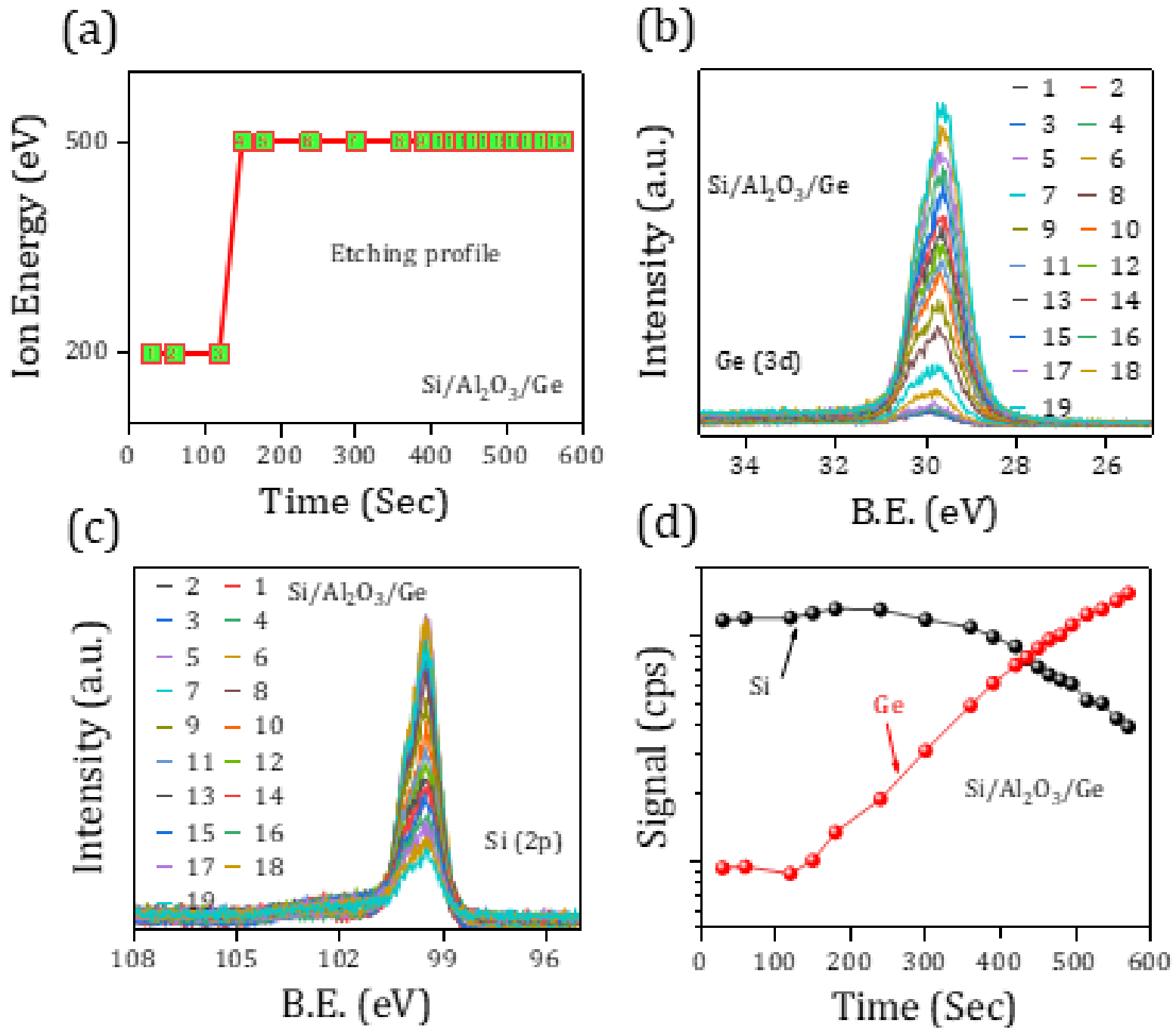

***Figure S5.*** *Etching profile data with each etching step for the $Si/Al_2O_3/Ge$ heterostructure: (a) etching profile, (b) Ge 3d, (c) Si 2p, and (d) change in signals of the Ge 3d and Si 2p with the etching time.*

Fig. S6 depicts the profile of XPS signal changes during stepwise in-situ $Ar^+$ ion etching employed to reach the buried $Si/Al_2O_3/Ge$ interface. Fig. S6(a) outlines the etching sequence as a function of ion energy and etching duration, showing a gradual reduction of the Si-NM thickness using two ion energies of 200 eV and 500 eV. Fig. S6(b) and S6(c) display the Ge 3d and Si 2p core-level spectra collected at successive etching steps. With increasing etching time, the Ge 3d signal increases systematically, while the Si 2p intensity decreases correspondingly. Fig. S6(d) shows the variation in Ge 3d and Si 2p peak intensities with etching time, illustrating the progressive exposure of the interfacial region.

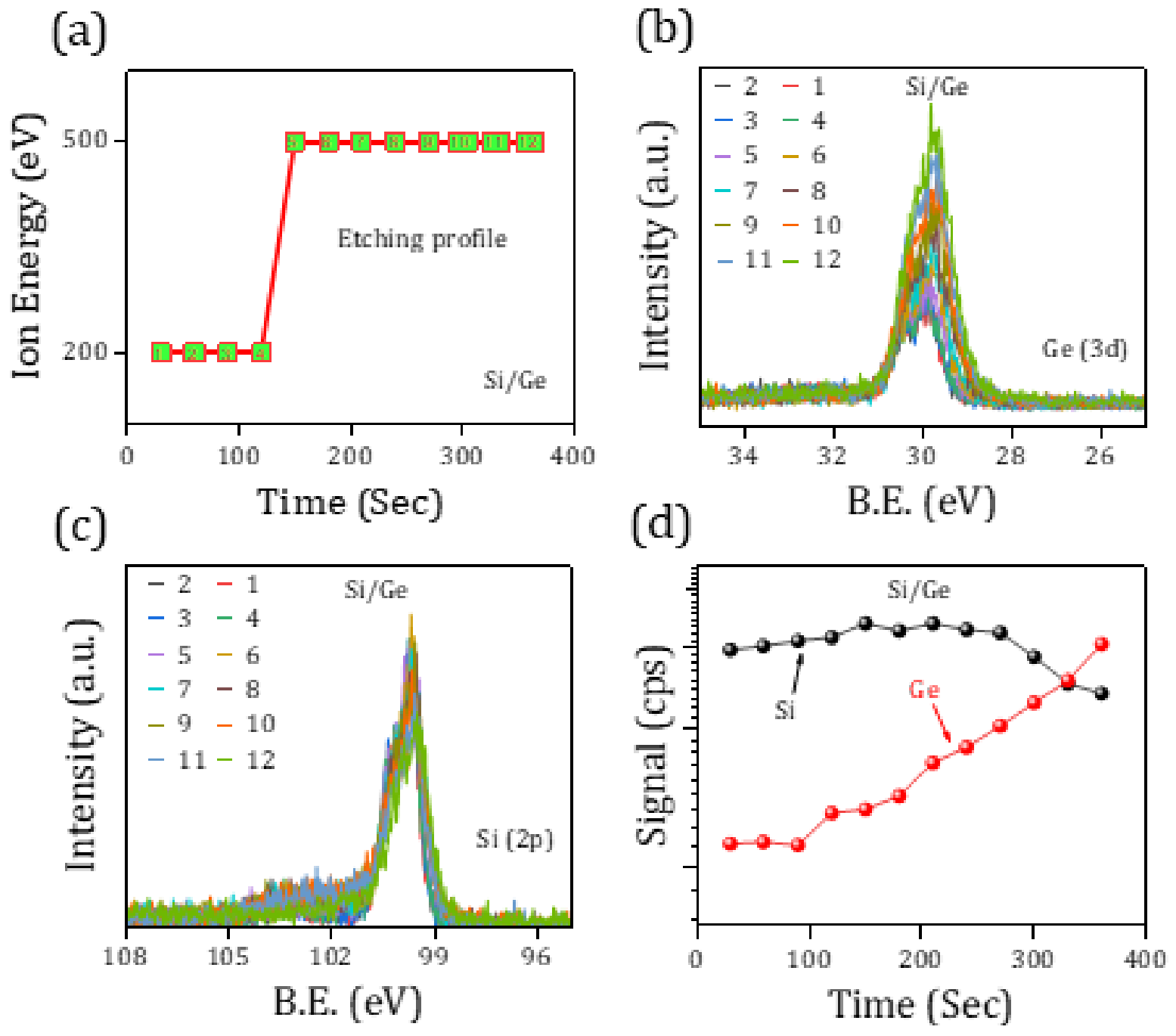

***Figure S6.*** *Etching profile data for each etching step for the Si/Ge heterostructure: (a) etching profile, (b) Ge 3d, (c) Si 2p, and (d) change in signals of the Ge 3d and Si 2p with the etching time.*

Fig. S7 shows high-resolution XPS core-level spectra of (a) C 1s, (b) O 1s, and (c) Al 2p acquired from the Si/$Al_2O_3$/Ge heterointerface after controlled ion etching. The C 1s spectrum exhibits a weak signal, indicating minimal carbon contamination at the interface. The O 1s spectrum confirms the presence of Al-O bonding associated with the ultrathin $Al_2O_3$ interlayer. At the same time, the Al 2p peak further verifies the presence of the $Al_2O_3$ layer at the bonded interface. These results confirm that the intended $Al_2O_3$ layer dominates the interfacial chemistry.

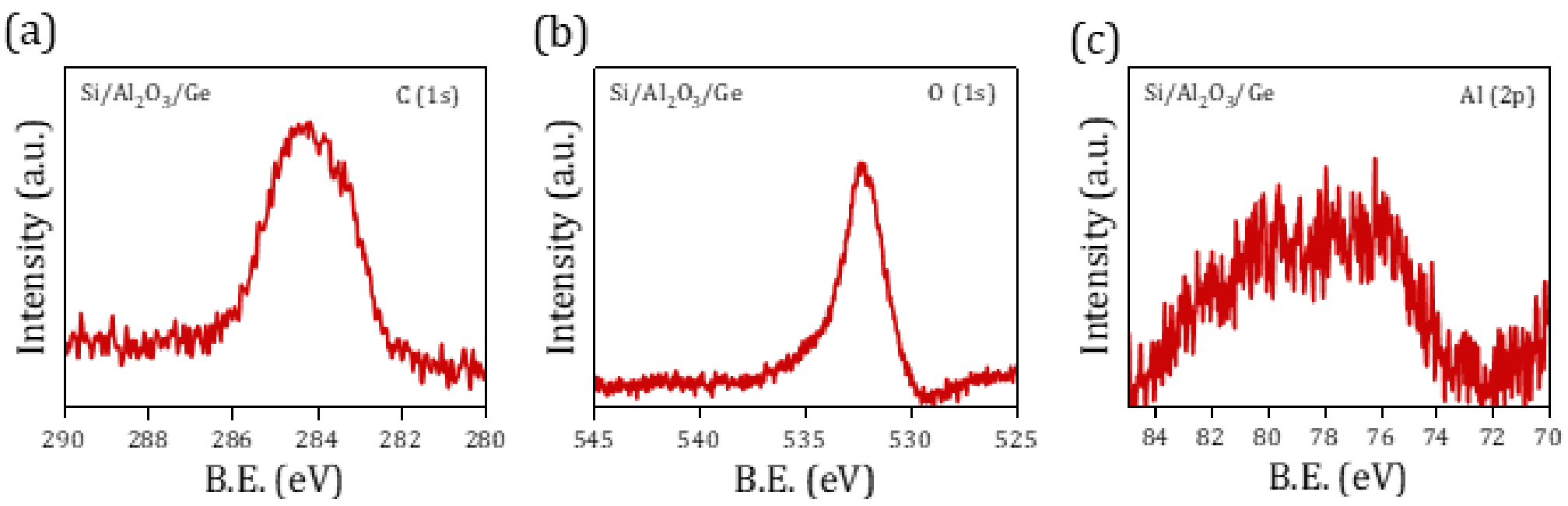

***Figure S7.*** *XPS core-level spectra of (a) C 1s, (b) O 1s, and (c) Al 2p at the interface of* $Si/Al_2O_3/Ge$ *heterostructure.*

High-resolution XPS core-level spectra acquired from the Si/Ge heterostructure are presented in Fig. S8. The C 1s and O 1s spectra shown in Fig. S8(a) and S8(b) indicate the absence of detectable surface carbon contamination and native oxide after the undercut of $SiO_2$ for the fabrication of a direct bonding heterostructure. In addition, the F 1s and Cl 2p spectra displayed in Fig. S8(c) and S8(d) show no residual fluorine or chlorine signals following the HF/HCl-based $SiO_2$ undercut procedure.

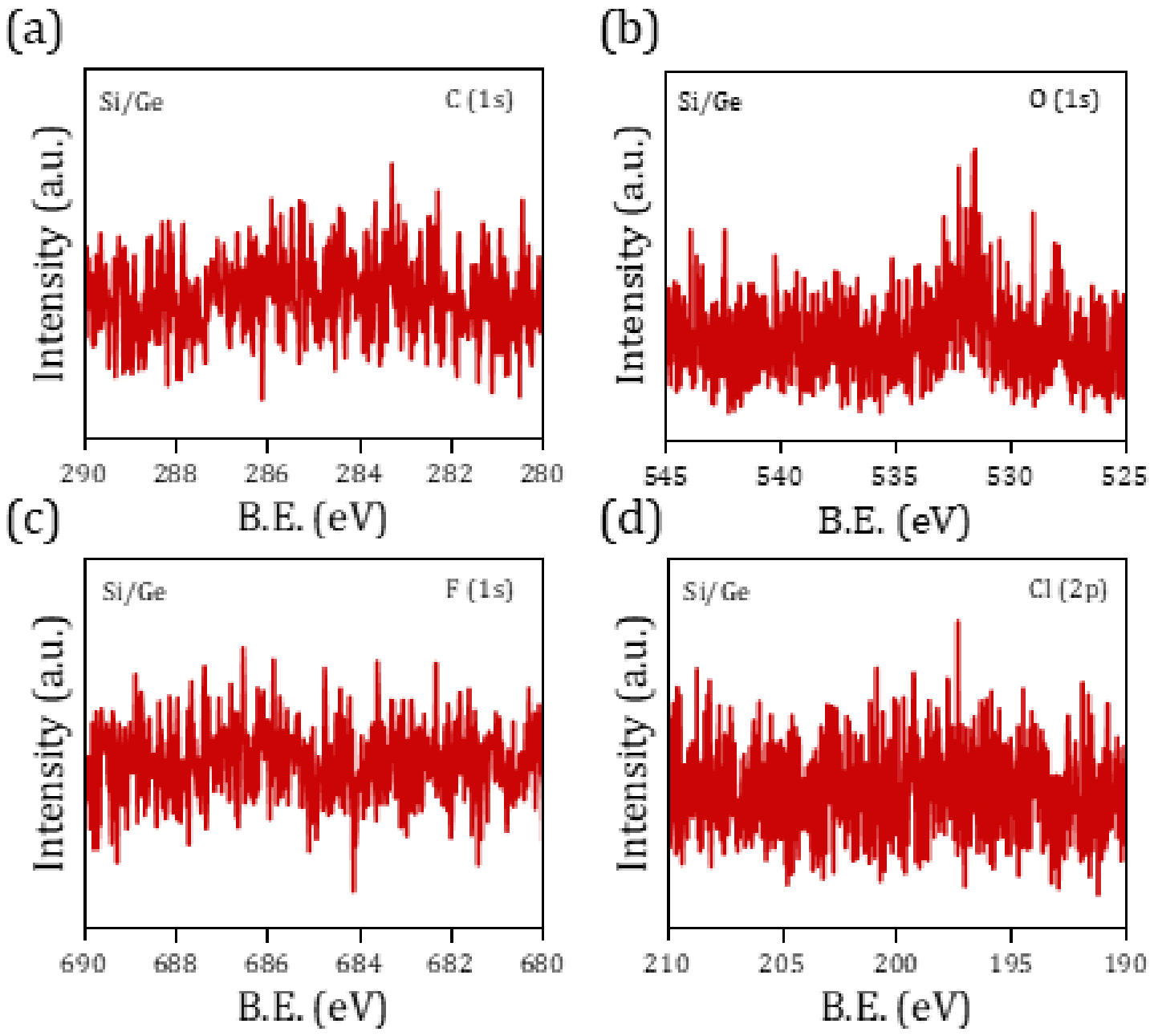

***Figure S8.*** *XPS core-level spectra of (a) C 1s, (b) O 1s, (c) F 1s, and (d) Cl 2p for the direct-bonded Si/Ge heterostructure.*

***Experimental Section***

Raman spectroscopy (Horiba LabRAM) equipped with a 633 nm excitation wavelength source was employed to investigate the vibrational characteristics of the samples. The structural quality and strain state of the Ge epilayer were assessed using HR-XRD (PANalytical Empyrean diffractometer equipped with a Cu $K\alpha_1$ source, $\lambda = 1.5406$ Å), while atomic force microscopy (AFM; Bruker) characterized the surface morphology. $Al_2O_3$ was deposited by ALD (Fiji G2, Cambridge Nanotech) at 200 °C, whereas $SiO_2$ was deposited using PECVD (Plasma-Therm 73/72) at 250 °C. Moreover, the ETCH Reactive Ion Etcher (Plasma-Therm 790) was utilized to etch the Silicon. XPS (Thermo Scientific K-alpha; calibration was performed using the Au $4f_{7/2}$ peak at 84.0 eV ) was used to analyze the chemical and electronic structure of the heterointerfaces.

# Refrences